Original Article:

# The Effects of Anthropogenic Air and Light Pollution on Astrophysics and Society

**Maya Nunez**

La Serna High School,

Whittier,

CA, 90605,

United States

**Correspondance**:

maya.nunez322@ gmail.com

Light and air pollution are the two main forms of pollution, containing the highest concentration in urban areas. I set out to investigate the effects that anthropogenic air and light pollution have on the night sky, how this affects the astronomical data-collecting process, and how the general public perceives both forms of pollution. This paper utilizes primary and secondary research, referring to different sources: data collected through my telescope, astrophotography, other scientists' research, and survey responses from over 40 people across 2 areas: Southern California and Guadalajara, Mexico. Through this, I have found strong correlations between increased aerosol particles from air pollutants and increased night sky brightness, increased aerosol particles amplifying cloud formation, which in turn increases light reflection, and a correlation between different areas and what they believe about pollution, with many being misinformed on air and light pollution themselves.

## 1.0  INTRODUCTION – AEROSOLS AND SKYGLOW

Light and air pollution are ceaseless and relentless, causing them to become the two main forms of pollution. Light pollution is most commonly caused by outdoor artificial lights pointing upward instead of at their intended target, with light pollution's most common manifestation being skyglow. Skyglow is created by combining artificial light with atmospheric scattering, producing a reduced contrast effect. The magnitude of skyglow is decided by aerosol particles in combination with the atmosphere's light processing effects, scattering this light at different magnitudes.

Light is comprised of photons, and these particle paths from its initial emitted light sources can change due to differences in aerosol particles. These distinctions result from differences in the size, nature, transportation, and concentration of particles in the atmosphere. The fewer aerosol particles that are accumulated in the atmosphere, the fewer photons that will scatter downward, and vice versa. Less accumulation of particles would mean a decrease in night sky brightness as observed from the ground, while an increase in particles would amplify night sky brightness. This effect is strongest near the light source, having the possibility of scattering its effects to over a 200-mile radius, as it has been noted in the National Park Service's article [1] on light pollution.

In order for aerosol particles to make it into the atmosphere, disturbances like wind currents must occur because the majority of aerosol particles settle to the ground from their initial site. Once gathered into the atmosphere, different light scattering phenomena are seen due to different factors. The AOD, or aerosol optical depth, is known to correlate well with the total concentration of particles in the air and is used to characterize the attenuation of light beams as they traverse the atmosphere. This also relates to the cross-section, or the probability of two particles colliding and acting a certain way [2].

Due to this, there is a diversity of optical effects of different amplitudes, caused by the mixture of aerosol particles in the atmosphere. Miroslav and Barentine [2] found that even for the same concentration of particles, photon scattering can vary over different magnitudes. These different magnitudes of scattering will differ in correspondence with the particle population's different sizes or chemical compositions. Even if the particles are of the same concentration, the particles can be comprised of different materials and are of random shapes and sizes, changing how the scattering will appear. The chemical composition and mean refractive index, or how fast a light beam travels through a medium compared to a vacuum, can be different for each collection of particle materials, as each collection is unique.

In general, Miroslav and Barentine's other findings show, that an increase of the particle size by 20 percent, 50 percent, and 100 percent over baseline while holding the number concentration constant enhances the scattered light intensity in all directions. Their findings also show that an increase in particle size and concentration are two root causes for how intense sky brightness is. The larger the particle or concentration of particles, the higher the amplitude of light scattering there is. They also found that air pollution control to reduce night sky brightness is strongest near a light source, meaning that light from sources at smaller distances contributes efficiently to the total brightness of the zenith sky -- the sky directly overhead -- because the relatively shorter optical paths of those light rays mean that they suffer less atmospheric attenuation from source to observer. LED lights, the newest and most adopted light source, are showing an increase in blue wavelength emissions, inducing significant negative impacts on visual astronomy. For this section, Kocifaj Miroslav and John C Barenstine's study "Air Pollution Mitigation Can Reduce the Brightness of the Night Sky in and Near Cities" [2] was used as my main source.

## 2.0 HOW CLOUD FORMATION RELATES TO AIR AND LIGHT POLLUTION

Cloud formation is a process done in the atmosphere that needs the contact of aerosol particles to proceed. Clouds play a major role in the brightness of skyglow, as clouds contain reflective properties. A NASA study led by scientist Jonathan Jiang collected data from 2006-2018 [3], using two satellites on the same track to show how clouds are formed in correspondence with human-made air pollution levels, revealing how detrimental air pollution can be to cloud formation.

   Clouds do not typically form without aerosols, as water vapor in the air does not easily condense without the contact of aerosol particles. There are many other aerosols other than anthropogenic ones, like volcanic ash and pollen, but anthropogenic pollution is much more incessant than these natural phenomena, causing larger impacts. As the first section states, aerosols come in different sizes, chemical compositions, concentrations, and transportation, affecting the way aerosols interact with clouds. Another additional factor with cloud formation and aerosol particles is the different effects at different altitudes in the atmosphere, which can cause the magnitude of these processes to differentiate.

   Natural aerosols, like smoke, absorb heat radiation emitted by the ground, warming the air through the smoke particles' increased temperatures. These particles also block sunlight, keeping the ground cooler, which overall reduces the temperature difference between the ground and the air. This narrows the temperature gap between the ground and the air, impeding cloud formation, since the ground needs to be cooler than the air for the moisture on the ground to evaporate and rise into the atmosphere.

   While natural aerosols absorb heat radiation and stop cloud formation, non-natural pollutants, like sulfates and nitrates, do not. When found in moderate concentrations, this adds more particles for cloud condensation into the atmosphere, feeding the process of cloud growth. Heavy pollution, on the other hand, shuts down cloud formation, as the number of particles in the atmosphere becomes too great. This blocks incoming sunlight and cools the ground, just as smoke aerosols do, inhibiting the formation of clouds.

   This extreme air pollution is only seen in the world's most polluted cities, and since most cities in the world are not that extreme, this increase in air pollution will only amplify cloud growth. This is harmful for Astronomy and other observers of the night sky, as, "Clouds over a city light dome can enhance the sky brightness by as much as 1000 times compared to a clear night sky" [1]. As this quote shows, not only do clouds block stars and other astrophysical objects, but the presence of clouds in the night sky increases sky brightness due to their light reflection properties, making it that much harder to view the stars.

## 3.0 MY DATA – INTRODUCTION

Living in Southern California, specifically a suburb of Los Angeles County that is only a 45-minute drive from the heart of Los Angeles, the effects of pollution have always been prominent. The particular city in Los Angeles County where my data was collected is La Mirada, which resides at an elevation of about 200 feet. Clear nights with no clouds here are few, and even on clear nights, the amount of stars in the sky is low. This is due to our night sky being so bright that it drowns out most stars, with the left side of my sky, facing towards Los Angeles, being much brighter than the right side of my night sky, giving me only one side of the sky to collect data from. My right side of the sky faces Orange County and the mountains farther down, containing a near-black color, as opposed to the left side's brighter grey. This is most likely due to the fact that Los Angeles releases more aerosol particles and photons than my area and the areas to my right do, causing the build-up of aerosol particles and increased sky brightness on the left side of my sky.

### 3.1 My Data – Astrophotography Findings

Knowing how bad pollution is in my area, I wondered how the effects of pollution differ in a more secluded, but still nearby, area. With this in mind, I decided to collect data while on my trip to Big Bear, California to compare the skies in both areas. Big Bear is located at an elevation of about 6,752 feet, compared to La Mirada's elevation of 200 feet, and is about 70 miles away from La Mirada. Even though Big Bear is one of the more touristy towns in the mountains, the difference in sky clarity was astonishing, as can be seen in Figure 1. There is a large difference in the clarity and brightness in both photos, with the photos from Big Bear coming out brighter and clearer. This illustrates how different elevations can affect how the atmosphere interacts with aerosol particles and how reduced distance from a major city can cause a reduction in night sky brightness.

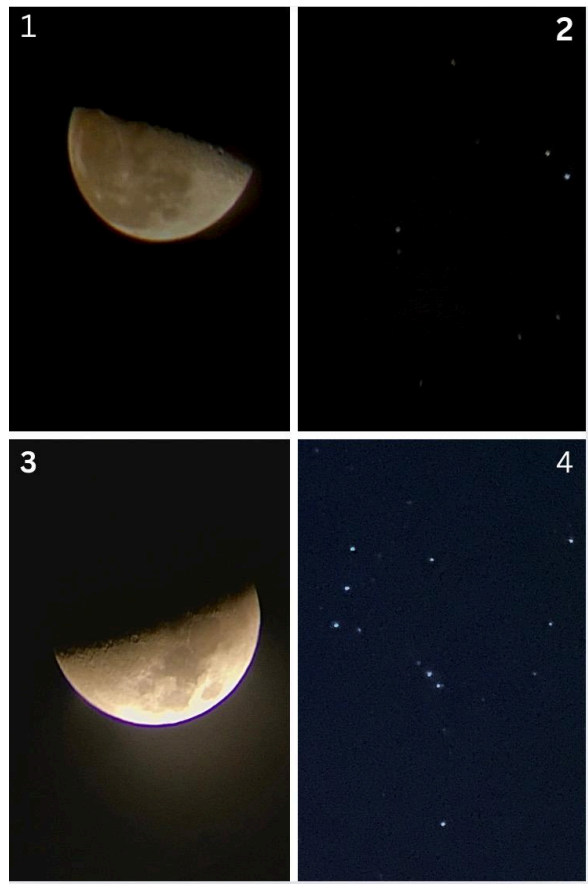

**FIGURE 1** Numbers 1 and 2 were taken in La Mirada, California, while Numbers 3 and 4 were taken in Big Bear, California. Numbers 1 and 3 are photos of the moon, while numbers 2 and 4 are of the constellation Orion. There is a large difference in the clarity and brightness of the photos, likely due to the elevation differences and reduced pollution seen in Big Bear, with it being much farther from any major city than La Mirada.

### 3.2 My Data – Survey Responses

Growing up in a city that neighbors one of the largest in the world, pollution was something that I was tone-deaf to until I visited my family in Northern California.

Being curious as to how educated other people are on the topic of anthropogenic air and light pollution and the effects in their area, I sent out a survey to 2 different areas: Southern California and Mexico. My group of respondents in California is 37, while my group in Mexico is much smaller, consisting of 4 people, and will be explained later. The questions and results for Southern California are attached below, with the results for Guadalajara, Mexico consisting of the same questions, just in Spanish, yielding different results.

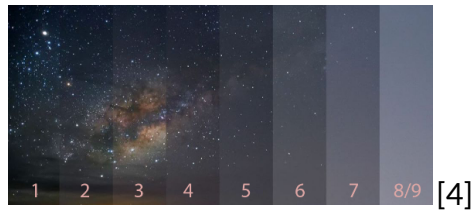 [4]

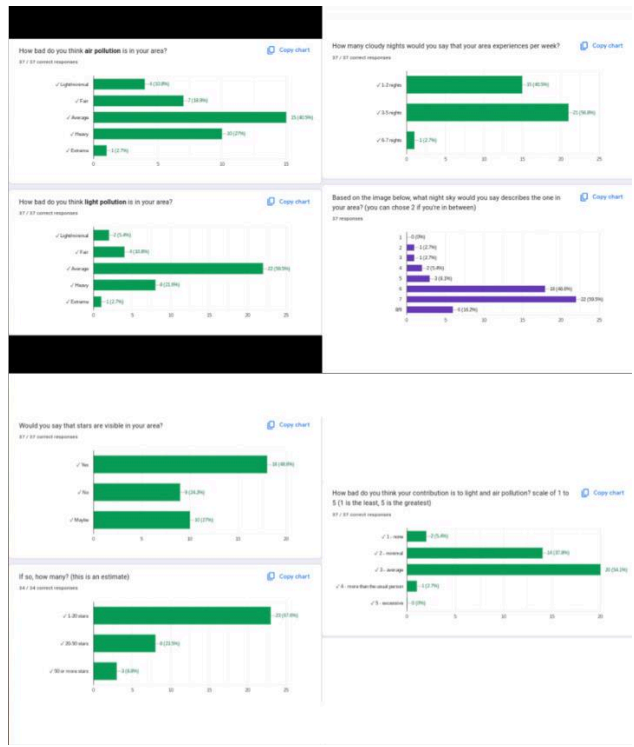

In the photos above, we see that for question 1, located at the top left, the largest response was that air pollution levels in Southern California are perceived to be average, with 40.5 percent, or 15 people, having this as their response. For question 2, we see, once again, "average" taking up the majority of the responses, meaning that 59.5 percent, or 22 out of 37 people, perceive their light pollution as average. For question number 3, located at the top left, we see that most people chose the middle response for 3-5 cloudy nights per week, with 56.9 percent, or 21 people, taking up this response. For question 4, they are responding to a night sky gradient, attached right above the responses. For the gradient, the top response was close, with 48.6 percent, or 18 people, responding 6, and 59.5 percent, or 22 people, responding 7. The next set of questions, 5 and 6, are at the bottom left and are a 2-part response. Question 5 asks whether stars are visible in your area, and these results are the closest yielded thus far. 48.6 percent, or 18 people, responded with yes, while 10 people responded maybe, and 9 people responded no, showing that the respondents possessed more uncertainty with this question. Question 6 had fewer responses due to uncertainty, with 67.6 percent, or 23 people, responding with 1-20 visible stars in the sky. Question 7, located on the bottom right of the photo, asked the responder to rate their contribution to pollution on a scale of 1-5, with once again, the average of 3 coming in with the most responses of 54.1 percent, or 20 out of 37 responses. Question 8 is the last one, not shown in the image as it is a short answer response. It asks the person who is filling out the survey to reflect on what daily tasks contribute to air and light pollution, and even though there was variation, 2 answers prevailed: driving a car and leaving your lights on at night. Even with these two responses, there were a select few people who answered that they are not sure what tasks contribute to this or that they have no contribution at all, which is concerning, and will be revisited after the analysis of Mexico.

For Guadalajara, Mexico, my control group was much smaller since I sent it to my dad's 4 siblings. Before getting into the results, for additional context, Guadalajara is the 4th largest city in Mexico based on population [5]. For all of my responses, I am translating them from Spanish to English. My results for question 1 are that 1 person responded average, 2 answered minimal, and 1 answered heavy, with minimal air pollution taking up 50 percent of the responses. The results for question 2 were 1 person for minimal, 2 for average, and 1 for extreme, with average light pollution being the main perception of Guadalajara at 50 percent. For question 3, the responses were 2 people for 1-2 nights and 2 people for 3-5 nights, sharing 50/50 results. For question 4, 3 people answered 5, while one answered 6, with most (75 percent of the responders) perceiving their night sky at the 5 mark. For question 5, the first part of the 2-part question, the results were that 3 responders claimed that stars are visible in their area, and 1 stated maybe, meaning that the majority perceive the stars as visible in Guadalajara, Mexico. The results of question 6, the follow-up to question 5, is that 1 person responded with 1-20 stars, 2 responded with 20-50 stars, and 1 responded with 50 plus visible stars; 50 percent perceive that their night sky holds 20 to 50 visible stars. For question 7, 3 people, or 75 percent, responded with minimal contributions to pollution, while 1 responded with more than the average person. For the answer to the last question, one trend was visible: using your car, with no mentions of light use.

## 3.3 Data Results – The Misinformation for the Public

With the results of section 3.1 -- one was the suburban area of Los Angeles, California, and the other was the urban area of Guadalajara, Mexico -- there were many misperceptions of anthropogenic air and light pollution in both groups. Starting with Southern California, many seemed to perceive the pollution as average in their area, but one Google search will tell you that Los Angeles County remains one of the most polluted regions in the nation [6]. Los Angeles County is where almost all of my responses were from, with a few being from neighboring cities, like La Habra or Yorba Linda, which are in Orange County and only a few miles away from La Mirada, making all responders share almost the same skies. Still, this one piece of information shows just how uninformed residents of Los Angeles County and other nearby areas are about air and light pollution. While the majority of people look toward the main city and say "While ours is less polluted, so it must not be that bad", many don't take into consideration that Los Angeles is one of the largest cities in the world and therefore, contains pollution levels far above the nation's average. Considering this, the slightly reduced pollution levels of the suburbs are nowhere near the nation's average, belonging in the heavy category as opposed to the average category, where only 27 percent of responses are found.

The second largest piece of misinformation is our contribution to anthropogenic pollution as residents. This was largely seen in the Southern California group, as I have realized through my yearly visits to Guadalajara, Mexico that their use of public transportation is much larger than here in Southern California. In Southern California, while emitted pollution levels may seem average, the tasks that contribute to air pollution go further than just the identified one of driving yourself to work or school, extending to tasks like using a refrigerator and eating beef. Many also recognize that leaving your light on contributes to light pollution, but many don't recognize that improper light fixtures that point in all directions and LED light usage also contribute to light pollution. Therefore, there is an overall underestimation when it comes to considering individual contributions to both forms of anthropogenic pollution. This is because many don't recognize that daily tasks all pile up into one big ecological footprint. There was also a select few who had no idea what contributed to pollution, and some believed that they had little to no contribution to pollution at all. This is concerning, as many don't realize that their ecological footprints cause accidental contributions to the pollution in their area, creating an atmosphere of unawareness.

Now that Southern California has been analyzed, the responses of Guadalajara, Mexico, must be looked over. The same trend of underestimation in pollution levels is shown, as Guadalajara, Mexico has air pollution levels above the average AQI or Air Quality Index. According to epa.gov, an AQI value of 50 or below represents good air quality, while an AQI over 300 represents hazardous air quality [7]. The AQI of Guadalajara on December 29 at 12:05 am was 152, or unhealthy, according to aqicn.org [8]. This means that the perceived level for the majority of the responses, fair air pollution, undermines the true pollution levels, exemplifying how growing up in a city can make you unaware of what bad pollution levels are. Light pollution was also overall perceived as average, but knowing how bright a city can get, this is also a misinterpretation. The fact that only car usage, and not light usage, was brought up as a main contributor to anthropogenic air and light pollution's effects on the night sky was astonishing, as skyglow cannot be amplified to that extent without the help of photons. This represents the same trend that was seen in Southern California, with residents of both urban areas containing an underestimation of pollution levels.

## 4.0 CONCLUSION – WHAT CAN BE DONE

With everything that has been stated in each section, the results from my data collection point towards two trends: large cities have major effects on neighboring suburban areas due to pollution and the residents of these areas have adapted to these unnatural levels, making it their norm. This normalcy towards pollution is something that many are tone-deaf to, as I have been guilty of it myself. It wasn't until I visited my family in Visalia, California, that I noticed that their night sky was the darkest I had ever seen, and that mine wasn't normal. This made me wake up to the unnaturally high air and light pollution levels in my area, making me realize that the scarcity of the stars is a result of our daily tasks as individuals. This is seen with the differences in my astrophotography, with Big Bear, California yielding brighter, more plentiful, stars.

Aerosols, which contribute to the trend above, are also a main contributor to air pollution. Aerosol particles are released through our daily tasks like driving a car and industrial processes, which in turn increases the scattering of light photons. The emissions of these photons are immense in cities and other urban areas, as the light fixtures in these areas are not concentrated on their object, releasing photons into the atmosphere. The combination of these factors above is what is causing the visibility of our night sky to wither at an astonishing rate, and urbanization will not help this. The underestimation of our contributions to pollution will also not help reduce these effects, as turning a blind eye to our daily tasks will never help fix the problem. Instead, we must understand that no matter what, we will always have an ecological footprint that will contribute to air and light pollution, but we must also realize that there are methods we can use to reduce the size of our impacts. We can carpool to school or work, walk to nearby places instead of driving, turn all unused lights off, and most importantly, use the correct outdoor light fixtures to have our lights only point towards our intended target and not up into the sky, releasing less photons into the atmosphere. If we do not do this, the future of astronomy, especially amateur astronomy in populated areas where pollution exists, is at risk, as the Astronomy field has already looked towards space telescopes as a method to combat this. My infatuation with the stars started at a young age, and knowing that anthropogenic air and light pollution can put the natural beauty of our universe at risk is concerning, as we are taking away future generations' ability to view the stars. Now, we must not allow our daily tasks to get in the way of the beauty of the stars, as we must allow our awareness of pollution to guide us toward a better future.

## ACKNOWLEDGEMENTS

A huge thank you to Murray Brightman of Caltech for being my Endorser and a pivotal person in helping me with the submission process. This paper couldn't have been submitted without his help.

**References**

[1] Service NP. Night Skies, Light Pollution. National Park Service (2024).

https://www.nps.gov/subjects/nightskies/lightpollution.htm

[2] Kocifaj M, Barentine JC. Air pollution mitigation can reduce the brightness of the night sky in and near cities. Scientific reports (2021). https://pmc.ncbi.nlm.nih.gov/articles/PMC8285390/

[3] NASA SET. NASA study untangles smoke, pollution effects on cloud growth. NASA (2018).https://science.nasa.gov/earth/earth-atmosphere/nasa-study-untangles-smoke-pollution-effects-on-clouds/

[4] Bidshahri R, The Effects of light pollution and the Bortle Scale. (2023)

https://rouzastro.com/astrophotography-a-guide-to-beating-light-pollution-part-1/

[5] Review WP. Mexico Cities by Population 2024. World Population Review (2024).

https://worldpopulationreview.com/cities/mexico

[6] Criteria Air Pollutants in Los Angeles County. County of Los Angeles Public Health. Accessed 28 December, 2024.

http://publichealth.lacounty.gov/eh/safety/criteria-air-pollutants.htm

[7] Air Data Basic Information – The AQI (Air Quality Index). United States Environmental Protection Agency (2024).

https://www.epa.gov/outdoor-air-quality-data/air-data-basic-information

[8] Guadalajara Air Pollution: Real-time Air Quality Index (AQI). https://aqicn.org/city/guadalajara/